\documentclass[letterpaper]{article} 
\usepackage{aaai2026}  
\usepackage{times}  
\usepackage{helvet}  
\usepackage{courier}  
\usepackage[hyphens]{url}  
\usepackage{graphicx} 
\usepackage{natbib}  
\usepackage{caption} 
\usepackage{algorithm}
\usepackage{algorithmic}

\usepackage{graphicx}
\usepackage{tabularx}
\usepackage{multirow}
\usepackage{booktabs} 
\usepackage{pgf-pie} 
\usepackage{pgfplots}
\usepackage{amsmath} 
\usepackage{pgfplots}
\pgfplotsset{compat=1.18} 
\usepackage{tikz}
\usepackage{xcolor}
\usepackage{tikz}
\usepackage{graphicx}   
\usepackage[table]{xcolor} 
\usepackage{diagbox}    
\usetikzlibrary{positioning} 
\usepackage{pifont}
\usepackage{amssymb}
\usepackage{multirow}
\usepackage{multicol}

\usepackage{newfloat}
\usepackage{listings}
\DeclareCaptionStyle{ruled}{labelfont=normalfont,labelsep=colon,strut=off} 
\floatstyle{ruled}
\newfloat{listing}{tb}{lst}{}
\floatname{listing}{Listing}
\title{FloodSQL-Bench: A Retrieval-Augmented Benchmark for Geospatially-Grounded Text-to-SQL}
\author {
    Hanzhou Liu,
    Kai Yin,
    Zhitong Chen,
    Chenyue Liu,
    Ali Mostafavi
}
\affiliations {
    Texas A\&M University\\
    \{hanzhou1996, kai\_yin, zhitong.chen18, liuchenyue\}@tamu.edu, amostafavi@civil.tamu.edu
}

\begin{document}

\maketitle

\begin{abstract}
Existing Text-to-SQL benchmarks primarily focus on single-table queries or limited joins in general-purpose domains, and thus fail to reflect the complexity of domain-specific, multi-table and geospatial reasoning,
To address this limitation, we introduce \textsc{FloodSQL-Bench}, a geospatially grounded benchmark for the flood management domain that integrates heterogeneous datasets through key-based, spatial, and hybrid joins.
The benchmark captures realistic flood-related information needs by combining social, infrastructural, and hazard data layers.
We systematically evaluate recent large language models with the same retrieval-augmented generation settings and measure their performance across difficulty tiers.
By providing a unified, open benchmark grounded in real-world disaster management data, \textsc{FloodSQL-Bench} establishes a practical testbed for advancing Text-to-SQL research in high-stakes application domains.
\end{abstract}

\begin{links}
    \link{Code}{https://github.com/HanzhouLiu/FloodSQL-Bench}
    \link{Datasets}{https://huggingface.co/datasets/HanzhouLiu/FloodSQL-Bench}
\end{links}

\section{Introduction}
Text-to-SQL translates natural language questions into executable SQL queries, enabling intuitive access to relational databases through conversational interfaces~\cite{katsogiannis2023survey}.
General-purpose benchmarks such as Spider~\cite{yu2018spider} and BIRD~\cite{li2023can} have significantly advanced research on this task.
However, these benchmarks primarily emphasize single-table queries or simple joins in open-domain settings, and thus fail to capture the challenges of domain-specific, multi-table and geospatial reasoning that are crucial for real-world applications.

Flood management is one such domain, where decision-making depends on integrating different data sources, including geographic features, demographic statistics, and critical infrastructure records~\cite{cutter2012social,wing2018estimates}. 
Unlike open-domain settings, practitioners must combine information across multiple layers, for example linking census tracts with floodplain boundaries or associating hospitals with historical flood claims. 
To overcome these challenges, we propose \textsc{FloodSQL-Bench}, a novel benchmark constructed from real-world datasets and designed around progressively complex queries, ranging from single-table lookups to multi-table reasoning involving key-based joins, spatial joins, and their hybrids. 
This design enables systematic evaluation of Text-to-SQL models on realistic geospatial and socio-economic reasoning tasks essential for flood risk assessment and emergency response.

We evaluate \textsc{FloodSQL-Bench} across a diverse set of large language models (LLMs), including both proprietary and open-source variants, under the same retrieval-augmented generation (RAG) settings. Beyond broad model coverage, we further compare model performance across the benchmark’s difficulty tiers to assess how well different LLMs handle increasing levels of schema complexity, table interactions, and geospatial operations.

This work makes the following contributions:
\begin{itemize}
\item We introduce \textsc{FloodSQL-Bench}, the first Text-to-SQL benchmark tailored to flood-risk analytics, featuring multi-table queries with geospatial reasoning.
\item We provide a comprehensive evaluation of multiple LLMs under RAG settings, systematically comparing their performance across difficulty levels ranging from single-table lookups to triple-table spatial–spatial joins.
\item Our study shows that while LLMs can handle simple queries reasonably well, they exhibit substantial degradation on complex multi-table and geospatial queries, highlighting the need for structured, metadata-driven, and domain-aware methods.
\end{itemize}


\section{Related Work}
\textbf{Text-to-SQL} research has evolved from single-table parsing to cross-domain and multi-table reasoning. Early datasets such as ATIS and GeoQuery focused on domain-specific grammars, whereas Spider~\cite{yu2018spider} established a large-scale cross-domain benchmark that has become the de facto standard for evaluating Text-to-SQL systems. Subsequent works, including SParC~\cite{yu2019sparc} and CoSQL~\cite{yu2019cosql}, extended Spider to multi-turn and conversational settings. More recently, benchmark efforts such as TableBench~\cite{wu2025tablebench} target broader table-based reasoning beyond SQL generation. However, all of these benchmarks primarily center on open-domain databases with relatively simple join structures, and rarely address domain-specific, multi-table reasoning that involves heterogeneous relational and spatial data, a key challenge for high-stakes applications such as flood risk management.

\medskip
\noindent
\textbf{Geospatial reasoning} introduces unique challenges beyond conventional relational semantics, as queries often involve spatial joins, coordinate transformations, and topology-aware aggregation.
Prior work on natural-language interfaces to spatial databases~\cite{li2019spatialnli, liu2025nalspatial} and spatial question answering~\cite{chen2021geoqa} has explored spatial relations such as containment, intersection, and proximity.
However, these systems are typically evaluated on synthetic or small-scale datasets and do not address realistic, domain-specific reasoning that integrates heterogeneous spatial and tabular information.
In contrast, real-world flood management requires the fusion of diverse geospatial sources—such as physical floodplains, demographic indicators, and critical infrastructure layers, to support actionable decision-making~\cite{cutter2012social, wing2018estimates}.
Despite their importance, no existing Text-to-SQL or QA benchmarks systematically incorporate such spatially grounded, multi-table reasoning tasks.
\textsc{FloodSQL-Bench} bridges this gap by providing a unified benchmark that combines key-based, spatial, and hybrid joins across real disaster-management datasets, enabling rigorous evaluation of large language models (LLMs) under retrieval-augmented (RAG) settings.

\medskip
\noindent
\textbf{Retrieval-Augmented Generation} (RAG) has become a central approach for grounding large language models (LLMs) in traditional NLP tasks like question answering and conversation~\cite{lewis2020retrieval, izacard2021leveraging, shuster2021retrieval}, with recent work extending retrieval beyond unstructured text to structured sources such as relational databases and tables~\cite{ayala2024reducing, wu2024stark}.
\
However, existing text-to-SQL benchmarks primarily focus on non-spatial schema and do not evaluate reasoning over geometric data, spatial joins, or multi-layer geospatial infrastructures. 
\
Although prior work has explored geospatial question answering using knowledge graphs \cite{li2025question} and GIS tool–driven workflow automation \cite{zhang2024geogpt}, neither line of research concentrates on SQL-based spatial joins across multiple tables. 
\
Recent advances in table-aware retrieval~\cite{zhang2023refsql, ziletti2024retrieval} further highlight the importance of retrieval granularity for structured reasoning, yet these methods have not been explicitly tested in spatial settings where geometry, topology, and coordinate systems must be considered. 
\
To the best of our knowledge, \textsc{FloodSQL-Bench} addresses these gaps by providing the first benchmark designed for multi-layer geospatial SQL reasoning and RAG evaluation, enabling systematic assessment of LLMs across lookup, relational, and spatially grounded analytical queries.
\section{Benchmark Construction}

\begin{table*}[t]
\centering
\setlength{\tabcolsep}{7pt}
\begin{tabular}{lcccccccccc}
\toprule
 & Tract & Flood & ZCTA & Schl & Hosp & Claim & Cnty & NRI & SVI & CRE \\
\midrule
Tract & \cellcolor{gray!30}— 
      & \cellcolor{blue!15}S 
      & \cellcolor{blue!15}S 
      & \cellcolor{blue!15}S 
      & \cellcolor{blue!15}S 
      & \cellcolor{orange!20}K 
      & \cellcolor{green!20}\textbf{S}/\textbf{K} 
      & \cellcolor{orange!20}K 
      & \cellcolor{orange!20}K 
      & \cellcolor{orange!20}K \\
Flood & \cellcolor{gray!10}\textcolor{gray}{S} 
      & \cellcolor{gray!30}— 
      & \cellcolor{blue!15}S 
      & \cellcolor{blue!15}S 
      & \cellcolor{blue!15}S 
      & \cellcolor{white}· 
      & \cellcolor{blue!15}S 
      & \cellcolor{white}· 
      & \cellcolor{white}· 
      & \cellcolor{white}· \\
ZCTA  & \cellcolor{gray!10}\textcolor{gray}{S} 
      & \cellcolor{gray!10}\textcolor{gray}{S} 
      & \cellcolor{gray!30}— 
      & \cellcolor{blue!15}S 
      & \cellcolor{blue!15}S 
      & \cellcolor{white}· 
      & \cellcolor{blue!15}S 
      & \cellcolor{white}· 
      & \cellcolor{white}· 
      & \cellcolor{white}· \\
Schl  & \cellcolor{gray!10}\textcolor{gray}{S} 
      & \cellcolor{gray!10}\textcolor{gray}{S} 
      & \cellcolor{gray!10}\textcolor{gray}{S} 
      & \cellcolor{gray!30}— 
      & \cellcolor{orange!20}K 
      & \cellcolor{white}· 
      & \cellcolor{blue!15}S 
      & \cellcolor{white}· 
      & \cellcolor{white}· 
      & \cellcolor{white}· \\
Hosp  & \cellcolor{gray!10}\textcolor{gray}{S} 
      & \cellcolor{gray!10}\textcolor{gray}{S} 
      & \cellcolor{gray!10}\textcolor{gray}{S} 
      & \cellcolor{gray!10}\textcolor{gray}{K} 
      & \cellcolor{gray!30}— 
      & \cellcolor{white}· 
      & \cellcolor{green!20}\textbf{S}/\textbf{K} 
      & \cellcolor{white}· 
      & \cellcolor{white}· 
      & \cellcolor{white}· \\
Claim & \cellcolor{gray!10}\textcolor{gray}{K} 
      & \cellcolor{white}· 
      & \cellcolor{white}· 
      & \cellcolor{white}· 
      & \cellcolor{white}· 
      & \cellcolor{gray!30}— 
      & \cellcolor{orange!20}K 
      & \cellcolor{orange!20}K 
      & \cellcolor{orange!20}K 
      & \cellcolor{orange!20}K \\
Cnty  & \cellcolor{gray!10}\textcolor{gray}{S/K} 
      & \cellcolor{gray!10}\textcolor{gray}{S} 
      & \cellcolor{gray!10}\textcolor{gray}{S} 
      & \cellcolor{gray!10}\textcolor{gray}{S} 
      & \cellcolor{gray!10}\textcolor{gray}{S/K} 
      & \cellcolor{gray!10}\textcolor{gray}{K} 
      & \cellcolor{gray!30}— 
      & \cellcolor{orange!20}K 
      & \cellcolor{orange!20}K 
      & \cellcolor{orange!20}K \\
NRI   & \cellcolor{gray!10}\textcolor{gray}{K} 
      & \cellcolor{white}· 
      & \cellcolor{white}· 
      & \cellcolor{white}· 
      & \cellcolor{white}· 
      & \cellcolor{gray!10}\textcolor{gray}{K} 
      & \cellcolor{gray!10}\textcolor{gray}{K} 
      & \cellcolor{gray!30}— 
      & \cellcolor{white}· 
      & \cellcolor{white}· \\
SVI   & \cellcolor{gray!10}\textcolor{gray}{K} 
      & \cellcolor{white}· 
      & \cellcolor{white}· 
      & \cellcolor{white}· 
      & \cellcolor{white}· 
      & \cellcolor{gray!10}\textcolor{gray}{K} 
      & \cellcolor{gray!10}\textcolor{gray}{K} 
      & \cellcolor{white}· 
      & \cellcolor{gray!30}— 
      & \cellcolor{white}· \\
CRE   & \cellcolor{gray!10}\textcolor{gray}{K} 
      & \cellcolor{white}· 
      & \cellcolor{white}· 
      & \cellcolor{white}· 
      & \cellcolor{white}· 
      & \cellcolor{gray!10}\textcolor{gray}{K} 
      & \cellcolor{gray!10}\textcolor{gray}{K} 
      & \cellcolor{white}· 
      & \cellcolor{white}· 
      & \cellcolor{gray!30}— \\
\bottomrule
\end{tabular}
\caption{
Join relationships among the ten tables in \textsc{FloodSQL-Bench}. 
\cellcolor{blue!15}\textbf{S} = spatial join; 
\cellcolor{orange!20}\textbf{K} = key-based join; 
\cellcolor{green!20}\textbf{S/K} = both spatial and key paths are supported; 
\cellcolor{gray!30}— self. 
The lower triangle mirrors the upper and is shaded in gray.
}
\label{tab:join-matrix}
\end{table*}

To balance realism with tractability, \textsc{FloodSQL-Bench} focuses on three flood-prone states, 
Texas, Florida, and Louisiana,
which together account for a disproportionate share of National Flood Insurance Program (NFIP) claims and Federal Emergency Management Agency (FEMA)–declared disasters. 
Within this scope, \textsc{FloodSQL-Bench} integrates ten heterogeneous tables across three spatial aspects: 
(i) non-spatial layers, \texttt{claims}, \texttt{svi}\footnote{\texttt{svi}: Social Vulnerability Index.}, \texttt{cre}\footnote{\texttt{cre}: Community Resilience Estimates.}, \texttt{nri}\footnote{\texttt{nri}: National Risk Index.}; 
(ii) polygon layers, \texttt{floodplain}, \texttt{census\_tracts}, \texttt{zcta}, \texttt{county}; and 
(iii) point layers, \texttt{schools}, \texttt{hospitals}. 
These layers reflect key information needed for flood risk management~\cite{cutter2012social, wing2018estimates}.

All external datasets used in the proposed benchmark~\textsc{FloodSQL-Bench} are sourced from publicly accessible U.S. government open-data portals, including datasets provided by~\citeauthor{census_data},~\citeauthor{fema_data},~\citeauthor{cdc_data}, and~\citeauthor{hifld_data}, 
encompassing demographic, infrastructure, hazard, and social-vulnerability data layers.

\subsection{Tabular Data Foundations}

\paragraph{Geographic Identifiers.}
\textsc{FloodSQL-Bench} adopts standardized geographic identifiers defined by the U.S. Census Bureau to ensure consistent key-based joins across heterogeneous tables~\cite{uscb_geoid}. 
At the tract level, an 11-digit \texttt{GEOID} encodes a 2-digit state code, a 3-digit county code, and a 6-digit tract code, serving as the primary join key among the \texttt{census\_tracts}, \texttt{claims}, \texttt{svi}, \texttt{nri}, and \texttt{cre} tables. 
At the county level, the 5-digit prefix of the tract \texttt{GEOID} uniquely identifies each county in the \texttt{county} and \texttt{hospitals} tables. 
At the ZIP level, the \texttt{ZIP} field serves as the geographic identifier for the \texttt{schools} and \texttt{hospitals} tables. 
Two spatial layers, \texttt{zcta} and \texttt{floodplain}, lack non-spatial join keys and are instead linked through spatial relationships only.

\paragraph{Spatial representation.}
\textsc{FloodSQL-Bench} standardizes spatial representations across various data sources. 
Specifically, polygon layers retain geometries to support polygon--polygon joins via spatial SQL functions such as \texttt{ST\_Intersects}, \texttt{ST\_Contains}, and etc.
In contrast, point layers store only explicit latitude and longitude (\texttt{LAT}\&\texttt{LON}) fields, 
with geometries constructed on demand using \texttt{ST\_Point(LON, LAT)}. 
To ensure efficiency, all geometries are projected to a common coordinate reference system (CRS)~\cite{goodchild1992geographical,burrough2015principles} and simplified with topology-preserving tolerances~\cite{visvalingam2017line,geos2023}.

\subsection{Join Semantics and Relations}

\textsc{FloodSQL-Bench} defines a unified set of join rules to ensure interpretable query semantics across all ten tables. 
All joins fall into two major categories, \emph{key-based joins} and \emph{spatial joins}. 
Key-based joins rely on standardized identifiers such as GEOID, COUNTYFIPS, and ZIP, enabling equality-based connections between non-spatial tables or between attribute tables and administrative layers. 
Specifically, tract-level 11-digit GEOIDs link \texttt{claims}, \texttt{census\_tracts}, \texttt{svi}, \texttt{nri}, and \texttt{cre}, while the 5-digit prefix of GEOIDs serves as county-level identifiers connecting to the \texttt{county} and \texttt{schools} tables. 
In addition, ZIP codes bridge \texttt{schools} and \texttt{hospitals}. 
Spatial joins, by contrast, operate on geometry relationships, following two subtypes, \emph{point--polygon} and \emph{polygon--polygon}. 
Point--polygon joins associate \texttt{schools} or \texttt{hospitals} with surrounding spatial layers (\texttt{census\_tracts}, \texttt{floodplain}, \texttt{zcta}, and \texttt{county}) through their \texttt{LAT}/\texttt{LON} coordinates. 
Polygon--polygon joins capture geometric overlaps among regional layers, including intersections between \texttt{floodplain}, \texttt{census\_tracts}, \texttt{zcta}, and \texttt{county}. 
Together, these 14 key-based and 14 spatial join rules form the core relational backbone of \textsc{FloodSQL-Bench}, enabling a wide range of key-based, spatial, and hybrid SQL queries. 
A summary of all join pairs is presented in Table~\ref{tab:join-matrix}.

\subsection{Metadata Builder}
\begin{figure*}[tb]
  \centering
  \includegraphics[width=0.96\linewidth, trim = 0.5cm 6.5cm 10.0cm 0.5cm]{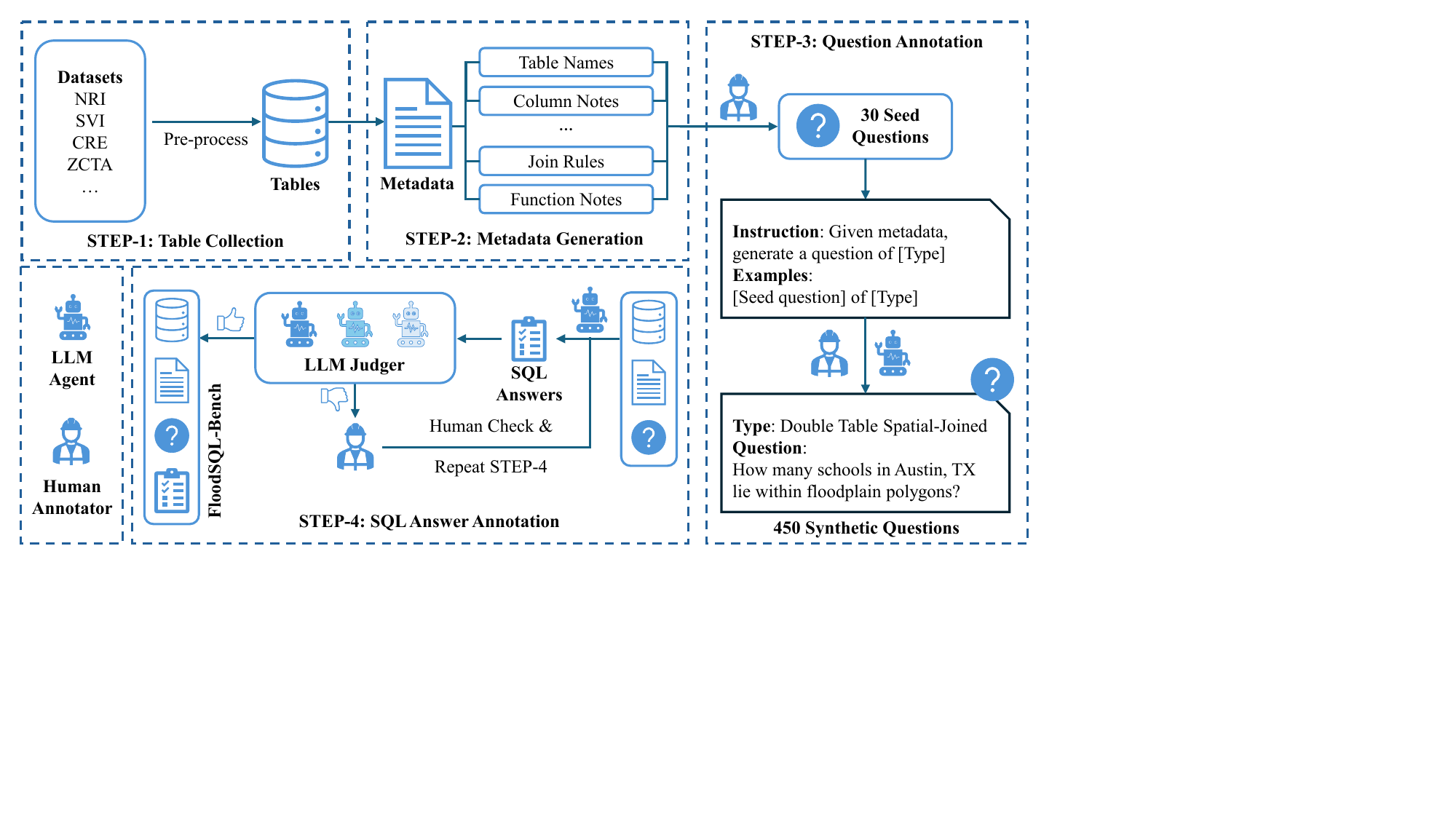}
  \caption{A simplified overview of the annotation framework of our proposed~\textsc{FloodSQL-Bench}.}\label{fig:annotation}
\end{figure*}
In a cross-table benchmark such as \textsc{FloodSQL-Bench}, 
metadata is not merely supplementary documentation but an integral component of the retrieval-augmented generation (RAG) framework itself. 
Without a structured metadata schema, it would be infeasible to construct a reliable RAG system capable of reasoning over heterogeneous tables. 
Accordingly, we design an enriched metadata schema that extends beyond basic table and attribute names, 
serving as the connective layer that bridges schema understanding and natural-language retrieval. 
It includes detailed field descriptions, data types, sample entries, join rules, function annotations, and supplementary notes, providing the semantic grounding necessary for accurate multi-table retrieval and reasoning.

\begin{table*}[tbp]
\centering
\renewcommand{\arraystretch}{1.25}
\resizebox{\textwidth}{!}{
\begin{tabular}{c|c|c|p{10cm}|p{8cm}}
\hline
\textbf{Level} & \textbf{\# of Tables} & \textbf{Join Types} & \textbf{Example Question} & \textbf{Sample Question Type} \\
\hline

\multirow[c]{4}{*}{\textbf{L0}} 
& \multirow[c]{4}{*}{1} 
& \multirow[c]{4}{*}{None} 
& 1) Return the fraction (0–1) of Texas census tracts where the estimated percent of population with no Internet access is greater than 20\%. 
& 1) fraction computation \\
&
&
& 2) Return the number of NFIP claims in Texas (identified by GEOID starting with 48) with zero insurance payout amount (in USD) for Increased Cost of Compliance (ICC). 
& 2) filter and count \\
&
&
& 3) Return the average non-null annual frequency of riverine flood events in Texas (identified by GEOID starting with 48). 
& 3) average computation \\
\hline

\multirow[c]{4}{*}{\textbf{L1}} 
& \multirow[c]{4}{*}{2} 
& \multirow[c]{4}{*}{Key} 
& 1)  Which non-null year had the highest total number of NFIP flood claims in Louisiana (STATEFP 22), based on a key-based join between claims and county tables? Return the year.
& 1) maximized aggregation and temporal ranking \\
&
&
& 2) Return the average non-null Normalized coastal flood risk score among unique Texas tracts (STATEFP 48) that have NFIP claims after 2000-01-01 and non-null NRI risk scores. 
& 2) average computation and multi-criteria filter \\
&
&
& 3) How many non-null ZIP codes in Florida (STATEFP 12) contain both at least one school and one hospital? 
& 3) intersection filter and distinct count \\
\hline

\multirow[c]{4}{*}{\textbf{L2}} 
& \multirow[c]{4}{*}{2} 
& \multirow[c]{4}{*}{Spatial} 
& 1) How many schools in San Antonio, TX (identified by CITY = 'SAN ANTONIO' and STATEFP = '48') fall inside floodplain polygons? 
& 1) point-in-polygon counting \\
&
&
& 2) Which census\_tract in Hillsborough County, FL (identified by STATEFP = '12' and COUNTYFP = '057') has the largest total overlap area with all zcta polygons? Return its 11-digit GEOID.
& 2) area-overlap ranking \\
&
&
& 3) How many pairs of census\_tracts in Hillsborough County, FL (STATEFP = '12' and COUNTYFP = '057') share a common boundary?
& 3) boundary-adjacency counting  \\
\hline

\multirow[c]{4}{*}{\textbf{L3}} 
& \multirow[c]{4}{*}{3} 
& \multirow[c]{4}{*}{Key-Key} 
& 1) Which 3 Texas counties (STATEFP 48) have the highest total non-null NFIP building payouts weighted by the non-null historical loss ratio for buildings due to coastal flooding? Return their county names.
& 1) weighted aggregation and ranking \\
&
&
& 2) In Harris County, TX (GEOID 48201), among tracts that have NFIP claims, what is the average non-null expected annual coastal flood loss per capita based on SVI total population? 
& 2) per-capita metric computation \\
&
&
& 3) List the 5 Texas (STATEFP 48) census tracts with the highest ratio of NRI riverine expected annual building loss to total housing units among tracts with NFIP claims. 
& 3) ratio computation and top-k selection \\
\hline

\multirow[c]{4}{*}{\textbf{L4}} 
& \multirow[c]{4}{*}{3} 
& \multirow[c]{4}{*}{Key-Spatial} 
& 1) For Louisiana (STATEFP 22), which 5 counties contain the largest number of schools located in floodplain areas? 
& 1) spatial-filtered counting \& ranking \\
&
&
& 2) For Florida (STATEFP 12), what is the average relative percentile rank for Summary percentile rank for Theme 1 across all census tracts that contain at least one hospital? 
& 2) spatial filter and attribute average computation \\
&
&
& 3) In Louisiana (STATEFP 22), what is the maximum Insurance payout amount (in USD) for structural building damage across all census tracts that contain at least one hospital? 
& 3) spatial filter and max aggregation \\
\hline

\multirow[c]{4}{*}{\textbf{L5}} 
& \multirow[c]{4}{*}{3} 
& \multirow[c]{4}{*}{Spatial-Spatial} 
& 1) How many hospitals are located within both FEMA floodplain polygons and census tract boundaries in Harris County, Texas (identified by the leftmost 5 digits of GEOID 48201 in the census\_tract table)?
& 1) multi-layer containment counting \\
&
&
& 2) What is the average total intersection area between each census tract and FEMA floodplain polygons, considering only tracts that also intersect ZCTA geometries, in Palm Beach County, Florida (identified by the leftmost 5 digits of GEOID 12099 in the census\_tract table)?  
& 2) multi-layer intersection-area aggregation \\
&
&
& 3) What percentage of census tracts in Dallas County, Texas (identified by the leftmost 5 digits of GEOID 48113 in the census\_tract table) intersect both FEMA floodplain polygons and county geometries? 
& 3) multi-criteria spatial percentage calculation \\
\hline

\end{tabular}
}
\caption{\textsc{FloodSQL-Bench} benchmark with example questions and their specific question types. The number of questions-SQL samples at each difficulty level, \{L0, L1, L2, L3, L4, L5\}=\{50, 100, 150, 50, 43, 50\}, resulting in a total number of 443 question-SQL pairs.}
\label{tab:floodsql_overview}
\end{table*}

\subsection{Question and SQL Annotation}
Figure~\ref{fig:annotation} illustrates the simplified progress of annotating \textsc{FloodSQL-Bench}.
Each question–SQL pair in \textsc{FloodSQL-Bench} is designed to reflect realistic analytical needs in flood management.
We construct natural language questions based on the underlying spatial and relational structure of the datasets, ensuring that each query is executable and grounded in real-world semantics.
All SQL queries are verified for correctness, and corresponding question texts are reviewed to balance linguistic diversity with structural clarity.
Table~\ref{tab:floodsql_overview} showcases sample questions with specific question types.

\paragraph{Question Annotation.} 
\textsc{FloodSQL-Bench} emphasizes diverse spatial and hybrid reasoning patterns that reflect real analytical workflows in flood risk assessment, rather than focusing solely on relational operations. 
To balance coverage and complexity, we organize 443 questions into six categories according to the number and type of joins involved, which are shown in Table~\ref{tab:floodsql_overview}.
This taxonomy captures the increasing difficulty from single-table loopups to triple-table geospatial reasoning, providing a structured foundation for evaluating model performance across progressively complex geospatial tasks.
Table~\ref{tab:floodsql_overview} reports the question categories, number of samples, description, and example questions in each category.

We follow the procedure of TableBench~\cite{wu2025tablebench} to construct diverse question–SQL pairs.
Specifically, we first manually design five seed questions for each category, covering representative reasoning patterns.
Next, we parse both the metadata and seed questions into the LLM agent, which automatically expand them into a larger set of candidate questions.
We then conduct a human review process to ensure quality and diversity, where annotators limit the frequency of specific tables, attributes, and operations (e.g., aggregation, area computation, \texttt{TOP-k} queries, “best” selection, or location comparison).
Finally, each question is manually rewritten for clarity and naturalness, ensuring that it remains human-readable and provides sufficient semantic hints for both human annotators and LLM agents to generate executable SQL.
Next, we describe the procedure used to construct the gold SQL answers in a fair and consistent way.

\paragraph{SQL Annotation.}
We employ an LLM agent to generate SQL queries under strict schema and function constraints defined in the metadata. To ensure consistency and interpretability, SQL generation is restricted to a limited yet well-defined set of functions, each aligned with specific reasoning categories. Inspired by TableBench~\cite{wu2025tablebench}, we incorporate a voting-based validation mechanism in which multiple LLM agents assess the correctness of each generated query. We iteratively refine each question–SQL pair until all three agents reach consensus that the SQL is valid and constitutes a reasonable answer to the question.
\
To further ensure quality, we execute every SQL query against the benchmark database to confirm that it runs successfully and produces a sensible output. Finally, although not illustrated in Figure~\ref{fig:annotation}, we perform an additional semantic consistency check in which human reviewers verify that each question faithfully reflects its corresponding SQL logic; any ambiguous or underspecified cases are manually revised for clarity.

\section{Experiments}
In this section, we evaluate recent LLM agents equipped with the same RAG framework and compare their performance in the flood-risk-analytics domain. Our benchmark spans various query types, including single-table reasoning, double-table key-based joins, double-table spatial joins, triple-table key–key joins, triple-table key–spatial joins, and triple-table spatial–spatial joins.

\begin{figure*}[tb]
  \centering
  \includegraphics[width=0.96\linewidth, trim = 1.1cm 9.7cm 15.5cm 0.7cm]{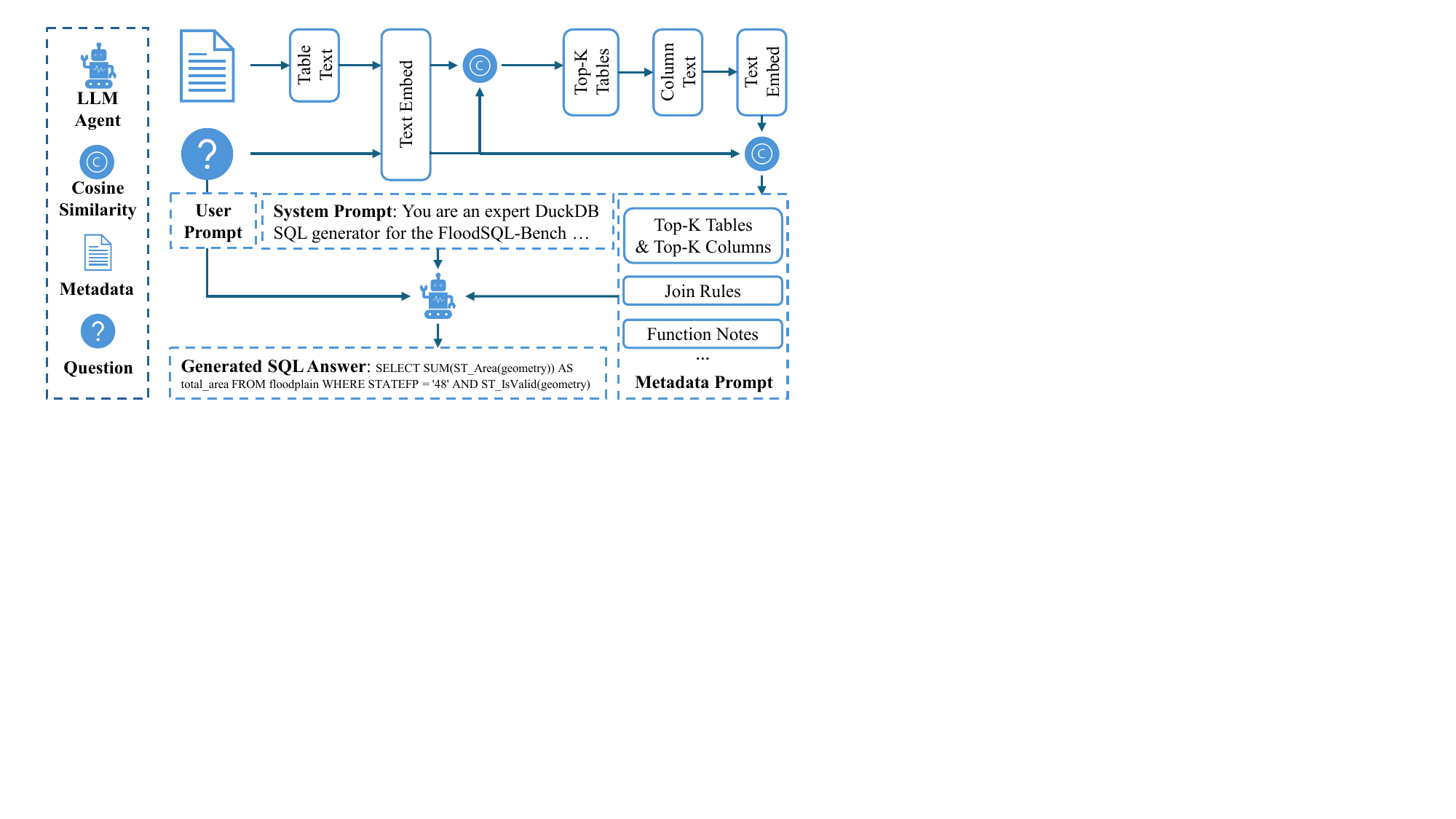}
  \caption{The Retrieval-Augmented Generation (RAG) architecture used to evaluate different LLM agents on \textsc{FloodSQL-Bench}. We compute text embeddings for both table-level and column-level metadata and measure their cosine similarity with the embedded question (user query) sequentially. The top-scoring tables and columns are selected as retrieved candidates for downstream SQL generation.}\label{fig:rag}
\end{figure*}
\subsection{Evaluation Protocol}
In geo-spatial text-to-SQL tasks, particularly those involving polygon layers, naively generated SQL queries often invoke expensive spatial operators, leading to long execution times. While human experts can manually design optimized SQL solutions that avoid such bottlenecks, LLM-based RAG systems frequently produce spatial joins or geometric operations that are too slow to execute in practice. A straightforward mitigation strategy is to impose a time threshold, but this approach is brittle because query latency varies significantly across hardware environments.
\
To address this issue, we propose a non-execution-based evaluation method: we embed both the ground-truth SQL and the LLM-generated SQL using OpenAI text-embedding-3-large and Jina Embeddings v3, and compute the cosine similarity between the resulting vectors. This avoids executing potentially expensive geospatial SQL, while still providing a reliable proxy for semantic similarity between candidate and reference queries.

\begin{table*}[t]
\centering
\small
\setlength{\tabcolsep}{2.2pt}
\begin{tabular}{lcccccccccccc}
\toprule
 & \multicolumn{2}{c}{\textbf{Level 0}} 
 & \multicolumn{2}{c}{\textbf{Level 1}} 
 & \multicolumn{2}{c}{\textbf{Level 2}} 
 & \multicolumn{2}{c}{\textbf{Level 3}} 
 & \multicolumn{2}{c}{\textbf{Level 4}} 
 & \multicolumn{2}{c}{\textbf{Level 5}} \\
\cmidrule(lr){2-3}
\cmidrule(lr){4-5}
\cmidrule(lr){6-7}
\cmidrule(lr){8-9}
\cmidrule(lr){10-11}
\cmidrule(lr){12-13}
\textbf{Model} &
\textbf{OpenAI} & \textbf{Jina} &
\textbf{OpenAI} & \textbf{Jina} &
\textbf{OpenAI} & \textbf{Jina} &
\textbf{OpenAI} & \textbf{Jina} &
\textbf{OpenAI} & \textbf{Jina} &
\textbf{OpenAI} & \textbf{Jina} \\
\midrule

\multicolumn{13}{l}{\textbf{Close-Source in Retrieval-Augmented SQL Generation Methods}} \\
\cmidrule(lr){1-13}
GPT-4o                  
& 0.914 & 0.962 & \textbf{0.891} & \textbf{0.957} & \underline{0.909} & \textbf{0.974} & 0.883 & \underline{0.951} & \textbf{0.904} & \textbf{0.962} & 0.919 & \textbf{0.976} \\
GPT-4.1                 
& 0.915 & 0.951 & 0.876 & 0.945 & 0.897 & 0.966 & 0.864 & 0.938 & 0.877 & 0.945 & 0.907 & 0.964 \\
GPT-5.1                 
& 0.917 & \underline{0.963} & 0.886 & \underline{0.953} & 0.899 & \underline{0.970} & 0.863 & 0.945 & 0.895 & 0.955 & 0.910 & 0.968 \\
Gemini-2.0-Flash-Lite$^*$   
& 0.897 & 0.934 & 0.869 & 0.945 & 0.907 & 0.962 & 0.885 & 0.948 & -- & -- & 0.914 & 0.964 \\
Gemini-2.5-Flash-Lite   
& 0.885 & 0.945 & 0.846 & 0.934 & 0.906 & 0.967 & 0.855 & 0.937 & 0.872 & 0.943 & 0.901 & 0.958 \\
Gemini-2.5-Pro          
& 0.887 & 0.943 & 0.872 & 0.952 & \textbf{0.914} & \textbf{0.974} & 0.862 & 0.940 & 0.883 & 0.955 & 0.923 & 0.970 \\
Claude-3-Haiku          
& \underline{0.919} & 0.957 & 0.862 & 0.939 & 0.895 & 0.960 & 0.870 & 0.947 & 0.895 & 0.957 & 0.895 & 0.960 \\
Claude-3.5-Haiku        
& 0.903 & 0.949 & 0.851 & 0.934 & 0.899 & 0.969 & 0.833 & 0.933 & \underline{0.898} & \underline{0.961} & 0.899 & 0.966 \\
Claude-4.5-Haiku$^*$        
& -- & -- & 0.870 & 0.938 & 0.896 & 0.966 & 0.879 & 0.949 & 0.895 & 0.958 & 0.921 & 0.973 \\
Claude-4.5-Sonnet         
& 0.907 & 0.950 & 0.856 & 0.931 & 0.904 & 0.969 & \underline{0.890} & \underline{0.951} & \textbf{0.904} & 0.952 & \underline{0.924} & 0.965 \\
Claude-4.5-Opus         
& \textbf{0.923} & \textbf{0.965} & \underline{0.889} & 0.951 & \underline{0.909} & \textbf{0.974} & \textbf{0.899} & \textbf{0.955} & -- & -- & \textbf{0.931} & \underline{0.974} \\

\midrule
\multicolumn{13}{l}{\textbf{Open-Source in Retrieval-Augmented SQL Generation Methods}} \\
\cmidrule(lr){1-13}
Deepseek-R1-7B$^*$         
& 0.732 & 0.766 & -- & -- & 0.752 & 0.812 & -- & -- & 0.776 & 0.843 & 0.777 & 0.812 \\
Deepseek-R1-14B$^*$          
& 0.802 & 0.868 & 0.792 & 0.888 & -- & -- & 0.765 & 0.853 & 0.803 & 0.882 & 0.826 & 0.886 \\
Deepseek-R1-32B$^*$          
& 0.848 & 0.903 & 0.843 & 0.925 & 0.887 & 0.961 & -- & -- & 0.889 & \underline{0.958} & 0.912 & \textbf{0.971} \\
Deepseek-V3.2-685B           
& \textbf{0.908} & \textbf{0.949} & \underline{0.873} & \textbf{0.949} & \textbf{0.908} & \underline{0.963} & \textbf{0.888} & \textbf{0.948} & \underline{0.896} & \underline{0.958} & \textbf{0.924} & \textbf{0.971} \\
Qwen3-1.7B               
& 0.861 & 0.916 & 0.827 & 0.910 & 0.846 & 0.924 & 0.793 & 0.886 & 0.824 & 0.900 & 0.856 & 0.926 \\
Qwen3-8B               
& 0.903 & 0.936 & 0.862 & 0.940 & \underline{0.904} & 0.961 & 0.871 & 0.942 & 0.875 & 0.946 & 0.912 & 0.967 \\
Qwen3-14B                 
& 0.895 & 0.933 & 0.864 & 0.940 & 0.893 & \underline{0.963} & 0.869 & 0.938 & 0.888 & 0.957 & 0.910 & \underline{0.970} \\
Qwen3-32B        
& 0.896 & 0.935 & 0.859 & 0.937 & 0.902 & 0.962 & \underline{0.873} & 0.943 & 0.893 & 0.957 & \underline{0.918} & \textbf{0.971} \\
Qwen3-235B-A22B       
& 0.894 & 0.942 & \textbf{0.877} & \underline{0.946} & 0.891 & \textbf{0.964} & 0.867 & \underline{0.947} & \textbf{0.898} & \textbf{0.962} & 0.916 & \textbf{0.971} \\
Gemma-2-2B$^*$          
& 0.879 & 0.931 & 0.820 & 0.907 & -- & -- & 0.841 & 0.919 & 0.818 & 0.909 & 0.851 & 0.924 \\
Gemma-2-9B          
& \underline{0.904} & \underline{0.944} & 0.865 & 0.942 & \underline{0.904} & 0.958 & 0.859 & 0.934 & 0.877 & 0.948 & 0.914 & 0.964 \\
\bottomrule
\end{tabular}

\caption{
Cosine similarity scores between predicted SQL results and gold SQL answers across the six benchmark task types:
single-table reasoning (level 0);
double-table key joins (level 1);
double-table spatial joins (level 2);
triple-table key--key joins (level 3);
triple-table key--spatial joins (level 4);
and triple-table spatial--spatial joins (level 5).
For each task type, we report scores computed with \textbf{OpenAI text-embedding-3-large} (OpenAI)
and \textbf{Jina Embeddings v3} (Jina).
All nine LLM--RAG agents are evaluated under the same RAG SQL generation framework as shown in Fig.~\ref{fig:rag}. The best results are \textbf{bold} while the second best results are \underline{underlined}. $^*$ marks LLM agents that exhibited timeouts or failure cases during SQL generation due to computation limitations. Following a best-effort validation process, any invalid or incomplete SQL outputs were removed from evaluation, denoted as --.
}
\label{tab:benchmark-types}
\end{table*}

\subsection{RAG Framework}
To support geospatial text-to-SQL generation in the flood-risk-analytics domain, we build a metadata-driven Retrieval-Augmented Generation (RAG) framework tailored to the structure of \textsc{FloodSQL-Bench}. As show in Fig.~\ref{fig:rag}, the system performs multi-granularity retrieval, first at the table level, then at the column level, before constructing a structured metadata prompt for SQL generation.

\paragraph{(1) Table catalog.}
Each relational or geospatial layer is represented by a table-level description that concatenates the table name with all schema fields and their summaries. During inference, the natural-language question is embedded and compared against these descriptions using cosine similarity. We select the Top-K most relevant tables (with K=3 for single-table tasks, K=4 for double-table tasks, and K=5 for more complex multi-table queries). This step constrains the LLM to operate only over valid tables and mitigates hallucination.

\paragraph{(2) Column-level descriptions.}
For each selected table, we maintain a column-level index where every entry pairs a column name with its semantic description. After table retrieval, the question is embedded again and matched against the column index of each chosen table. We then select the Top-5 most relevant columns per table, providing the LLM with fine-grained grounding (e.g., tract identifiers, FIPS codes, polygon geometries). These retrieved entries help the model map linguistic cues in the question to the appropriate attributes.

\paragraph{(3) Functional notes.}
We include a global set of operational notes summarizing CRS conventions, geometry validity requirements, common preprocessing steps, and warnings about expensive spatial operations. These notes bias the model toward efficient and domain-appropriate SQL patterns, preventing common spatial-analytics errors.

\paragraph{(4) Join rules.}
Because \textsc{FloodSQL-Bench} spans heterogeneous data sources, we provide explicit join rules covering direct key-based joins, concatenated ID joins, polygon-polygon spatial intersections, and point-in-polygon containment relationships. Exposing these rules gives the model structural priors that greatly reduce invalid joins and guide it toward plausible execution plans.

\paragraph{Inference workflow.}
Given a question, the system performs (i) dense table retrieval (Top-K), (ii) dense column retrieval for each selected table (Top-5), and (iii) metadata assembly that merges retrieved tables, selected attributes, functional notes, and join rules into a structured context. This enriched prompt is fed to the LLM, which is instructed to output exactly one DuckDB SQL query using only the retrieved tables and columns. This metadata-aligned RAG pipeline reduces ambiguity, improves schema grounding, and enhances the efficiency of generated geospatial SQL.
\textbf{We use \textsc{null} to denote the failed SQL predictions}.

\subsection{Results and Analysis}
Table~\ref{tab:benchmark-types} summarizes the performance of more than twenty state-of-the-art LLMs equipped with the same retrieval-augmented (RAG) SQL generation pipelines on \textsc{FloodSQL-Bench}. Across all models, we observe that the proposed RAG framework, supported by metadata such as join rules, general notes, and function descriptions, consistently enables strong SQL generation performance, indicating the effectiveness of structured retrieval guidance for cross-table geospatial reasoning.

Among all closed-source LLM agents evaluated, Table~\ref{tab:benchmark-types} shows that GPT-4o and Claude-4.5-Opus demonstrates the strongest overall performance across all difficulty levels. More specifically, we observe that recent GPT models, including GPT-4o, GPT-4.1, and GPT-5.1, exhibit consistent and comparable SQL generation performance on \textsc{FloodSQL-Bench}. A similar pattern holds for the Gemini family and Claude series. For example, models ranging from Gemini-2.0-Flash-Lite to Gemini-2.5-Pro show relatively stable performance across benchmark categories, and Claude models from Claude-3-Haiku to Claude-4.5-Opus likewise achieve closely aligned results.

Among all open-source LLM agents evaluated, DeepSeek-V3.2-685B demonstrates the strongest overall SQL generation capability under our proposed RAG framework. Within the DeepSeek family, DeepSeek-V3.2 achieves the highest overall performance across all categories. Other DeepSeek variants follow a clear scaling trend in which larger model sizes yield more accurate SQL predictions.
\section{Conclusion}
In this work, we presented \textsc{FloodSQL-Bench}, the first Text-to-SQL benchmark specifically designed for flood-risk analytics and multi-layer geospatial reasoning. By integrating real-world datasets spanning demographic, infrastructural, and floodplain information, the benchmark introduces progressively complex query types that require models to perform key-based joins, spatial joins, and hybrid multi-table reasoning. Our comprehensive evaluation across a broad set of proprietary and open-source LLMs, under a unified RAG framework, reveals that with appropriate retrieval settings and metadata hints, model performance on more challenging spatial and multi-table queries can approach that of simple single-table tasks.
\
\textsc{FloodSQL-Bench} enables more robust evaluation, fosters the development of specialized methods, and ultimately supports more reliable decision-making in flood management and related geospatial domains. Our benchmark will be open-sourced.

\section{Acknowledgments}
This work used the DeltaAI system at the National Center for Supercomputing Applications [award OAC 2320345] through allocation [allocation number CIV250031] from the Advanced Cyberinfrastructure Coordination Ecosystem: Services \& Support (ACCESS) program, which is supported by National Science Foundation grants \#2138259, \#2138286, \#2138307, \#2137603, and \#2138296.

\bibliography{aaai2026}

\end{document}